\documentclass[preprint,prb,aps]{revtex4}
\usepackage{color}
\makeatletter

\usepackage{amsmath}
\usepackage{graphicx}
\usepackage{bm}
\usepackage{fancyhdr}
\usepackage{amsmath}
\usepackage{graphicx}
\usepackage{amsfonts}
\usepackage{amsbsy}
\usepackage{amssymb}
\usepackage[mathscr]{eucal}

\usepackage{color}

\graphicspath{{./figures/}}

\usepackage{graphicx}
\usepackage{bm}

\usepackage{amsfonts}
\usepackage{amsbsy}
\usepackage{amssymb}
\usepackage[mathscr]{eucal}

\newcommand{\beq}{\begin{equation}}
\newcommand{\eeq}{\end{equation}}
\newcommand{\ba}{\begin{array}}
\newcommand{\ea}{\end{array}}
\newcommand{\bea}{\begin{eqnarray}}
\newcommand{\eea}{\end{eqnarray}}
\newcommand{\bseq}{\begin{subequations}}
\newcommand{\eseq}{\end{subequations}}

\date{today}

\begin{document}

\title{The Electron-Phonon Interaction of Low-Dimensional and Multi-Dimensional  Materials from He Atom Scattering}

\author{G. Benedek$^{a,b}$, J. R. Manson$^{a,c}$, S. Miret-Art\'es$^{a,d}$}
\email{giorgio.benedek@unimib.it, jmanson@clemson.edu,s.miret@iff.csic.es}
\affiliation{$^a$ Donostia International Physics Center (DIPC),~Paseo Manuel de Lardiz{{a}}bal,~4,~20018 Donostia-San Sebastian, Spain}
\affiliation{$^b$ Dipartimento di Scienza dei Materiali,~Universit{\`a} di Milano-Bicocca,~Via Cozzi 55,~20125 Milano, Italy}
\affiliation{$^c$ Department of Physics and Astronomy, Clemson University, Clemson, South Carolina 29634, U.S.A.}
\affiliation{$^d$Instituto de F\'isica Fundamental, Consejo Superior de Investigaciones
Cient\'ificas, \\ Serrano 123, 28006 Madrid, Spain \\}

\date{\today}

\begin{abstract}
	
Atom scattering is becoming recognized as a sensitive probe of the electron-phonon interaction parameter $\lambda$ at metal 
and metal-overlayer surfaces. Here, the theory is developed linking $\lambda$ to the thermal attenuation of atom scattering spectra 
(in particular, the Debye-Waller factor), to conducting materials of different dimensions, from quasi-one dimensional systems such 
as W(110):H(1$\times$1) and Bi(114), to quasi-two dimensional layered chalcogenides and high-dimensional surfaces such as 
quasicrystalline 2ML-Ba(0001)/Cu(001) and d-AlNiCo(00001). Values of $\lambda$ obtained using He atoms compare favorably with 
known values for the bulk materials. The corresponding analysis indicates in addition the number of layers  contributing to the 
electron-phonon interaction that is measured in an atom surface collision.

\end{abstract}

\maketitle

\vspace{2cm}


\newpage

\section{Introduction}

Electron-phonon (e-ph) interaction at conducting surfaces together with its dimensionality are of great importance both at fundamental and technological levels. \cite{Saito-17,WZhang-18} Very recently, this interaction has been shown to play a relevant role in topological semimetal surfaces  such the quasi-one-dimensional charge density wave system Bi(114) and the layered pnictogen chalcogenides. \cite{JPCL-2020}
The e-ph coupling  in these materials for individual phonons $\lambda_{{\bf Q}, \nu}$ (where ${\bf Q}$ denotes the surface parallel wave vector and $\nu$ the branch number), and its average $\lambda$, the well-known  mass-enhancement factor \cite{McMillan-68,Grimvall,Allen} can be directly measured with supersonic He-atom scattering (HAS).\cite{Skl,Benedek-14,Manson-JPCL-16,JPCL2,Manson-SurfSciRep} With this experimental technique, subsurface phonons were detected on multilayer metallic structures\cite{Skl,JPCL2} exploring the fairly long range of the e-ph interaction, e.g., spanning as many as 10 atomic layers in Pb films\cite{Skl,Benedek-14} (known as the {\em quantum sonar effect}). Under reasonable approximations, from the thermal attenuation of the diffraction peaks ruled by the so-called Debye-Waller (DW) factor as well as the interaction range through the number of layers $n_{sat}$, $\lambda_{HAS}$ values can actually be extracted which agree fairly well with previous values for the bulk or obtained from other surface techniques.\cite{JPCL-2020,Manson-JPCL-16,JPCL2} This quantity $n_{sat}$ indicates the number of layers above which the measured $\lambda$ becomes thickness-independent. 

In this work we focus on the specific role of dimensionality in the e-ph mass-enhancement factor as derived from HAS.\cite{JPCL-2020} In particular, we extract $\lambda_{HAS}$ values from HAS data for different classes of conducting surfaces characterized by nearly free-electron gases of growing dimensions, from the quasi-1D systems such as W(110):(1$\times$1)H and Bi(114), and the quasi-2D 
layered chalcogenides, to quasicrystalline surfaces such as the dodecagonal 2ML-Ba(0001)/Cu(001) and decagonal d-AlNiCo(00001), which can be regarded as behaving like  periodic 4D and 5D materials, respectively.

\section{Theory}

\subsection{The new Debye-Waller factor} 

As is well known, the DW factor describes the thermal attenuation, due to the surface atomic motion, of the elastic scattered intensity $I(T)$ observed at temperature $T$ with respect to the elastic intensity of the corresponding rigid surface $I_0$. It is usually written as 
\begin{eqnarray} \label{Eq1-2}
I(T) ~=~ I_0 e^{-2W(T)}   
~,
\end{eqnarray}
where the factor, $\exp\{ -2 W({\bf k}_f, {\bf k}_i ,T)  \}$, depends explicitly on the final (${\bf k}_f$) and incident (${\bf k}_i$) wave vectors of the scattered atom.
When zero point motion vibrations can be neglected, which holds for $T$ comparable to or larger than the surface Debye temperature of the material, $2W(T)$ is approximately linear in $T$. 

When the incident atom directly interacts with the surface, the DW exponent is simply expressed by
$2 W({\bf k}_f, {\bf k}_i ,T) = \left\langle (\Delta {\bf k}\cdot {\bf u} )^2  \right\rangle_T$,
$\Delta {\bf k} = {\bf k}_f -{\bf k}_i$ being the scattering vector, ${\bf u}$ the phonon displacement experienced by the projectile atom upon collision, and $ \left\langle \cdot \cdot \cdot \right\rangle_T$ means a thermal average. For energies typically well below 100 meV, incident atoms are exclusively scattered by the surface free-electron density, a few~\AA~away from the first atomic layer. Thus, the exchange of energy with the phonon gas mainly occurs via the phonon-induced modulation of the surface free-electron gas or the so-called e-ph interaction. 
Under reasonable approximations, it has been recently shown that the DW exponent is proportional to $\lambda$ and in the simplest case reads as\cite{Manson-JPCL-16}
\begin{eqnarray} \label{Eq1}
2W({\bf k}_f, {\bf k}_i,T) ~\cong~ 4 \mathcal{N}(E_F) ~ \frac{m E_{iz}}{m_e^* \phi} ~ \lambda ~ k_B T
~,
\end{eqnarray}
where $\mathcal{N}(E_F)$ is the electron density of states at the Fermi energy $E_F$, $m$ and $m_e^*$ are the projectile atomic mass and the electron effective mass, respectively, $\phi$ is the work function and $k_B$ is the Boltzmann constant.
Eq.~(\ref{Eq1}) is written here specifically for the specular diffraction peak for which
$E_{iz} = E_i \cos^2(\theta_i) = \hbar^2 k_{iz}^2/2m$ is the incident energy associated with the motion normal to the surface for the given incident angle $\theta_i$.  For application to non-specular diffraction peaks or to other elastic features, Eq.~(\ref{Eq1}) needs to be adjusted to account for the correct scattering vector appropriate to the scattering configuration, i.e., $4 k_{iz}^2 \longrightarrow \Delta {\bf k}^2 = ({\bf k}_f - {\bf k}_i)^2$ as discussed below.
Using the simple expression of Eq.~(\ref{Eq1}), a previous analysis of the thermal attenuation of the specular peaks of
He atom scattering from
several simple metals extracted values of $\lambda_{HAS}$. \cite{Manson-JPCL-16}  These values  must be regarded as values of the electron-phonon constant relevant to the region near the surface.  

An important observation concerning the DW factor $2W({\bf k}_f, {\bf k}_i,T)$ is that it is rigorously proportional to the temperature, subject to the condition that  $T$ is large compared to the Debye temperature (which in practice means that  $T$ should be larger than about half the Debye temperature).  This results because the average phonon mean-square displacement satisfies similar conditions as long as the crystal obeys the harmonic approximation.  Thus, plots of $2W({\bf k}_f, {\bf k}_i,T)$ versus $T$ (usually called DW plots) have linear slopes at large $T$, although at small values of $T$ where zero point motion becomes effective the curve saturates to a constant value.

There have been extensive He atom scattering measurements of successive layers of alkali metals grown on various metal substrates, and similar studies of multiple Pb layers grown on Cu(111).\cite{JPCL2}  These systems are of interest for studies of the e-ph constants, for example to see how $\lambda_{HAS}$ varies as a function of numbers of monolayers. The case of multiple Pb monolayers is of particular interest because thin films of Pb on Si and GaAs are known to remain superconductors down to one monolayer.
\cite{Zhang-2010,Uchihashi-2011,Sekihara-2013,Skl,Benedek-2018} An interesting first observation of the thermal attenuation of specular He atom scattering for layer-by-layer growth  is that, for a given system, the slope of the DW plots increased linearly with layer number $n$ for the first few layers up to a saturation number $n_{sat}$ which was typically about five layers.\cite{JPCL2}
This behavior suggests that each layer of these simple metals  contributes similarly and additively to the Fermi level density of states appearing in Eq.~(\ref{Eq1}), a property
which is also indicated by theoretical calculations for multi-layer alkali films.\cite{overlayers}
For the surface of bulk simple metals, using for the Fermi level density of states that of a three-dimensional (3D) free electron gas,
$\mathcal{N}(E_F) = 3 Z  m_e^*/ \hbar^2 k_F^2$, where $k_F$
is the Fermi wave vector and $Z$ the number of free electrons per atom, was a satisfactory approximation. \cite{JPCL2} 
However, the linearly increasing slope  observed for ultrathin films with $n < n_{sat}$ suggests that it is appropriate to attribute to each metal layer an independent contribution of the 2D free electron density of states (DOS), which implies that
$\mathcal{N}(E_F) = n  m_e^* a_c / \pi \hbar^2$ for $n \leq n_{sat}$ where $a_c$ is the area of the surface unit cell.
Combining this last expression (2D form) for the Fermi level density of states, and recognizing that $\ln[I(T)/I_0]= - 2 W$, the following expression for $\lambda$ from Eq.~(\ref{Eq1}) is obtained
\begin{eqnarray} \label{alpha}
\lambda_{HAS} ~=~ \frac{\pi}{2n} \alpha  ~~~; ~~~\alpha ~\equiv~ \frac{\phi~ \ln[I(T_1)/I(T_2)]}{a_{c} k^2_{iz} k_B (T_2-T_1)}~ ,   ~~~\mbox{for}~~n \leq n_{sat} ~,
\end{eqnarray}
where $T_1$ and $T_2$ are any two temperatures in the linear region of the DW plot.
If $n_{sat}$ is known one can use the 2D expression for $n > n_{sat}$ by setting $n=n_{sat}$ (this is what has been done for all layered crystals); alternatively, the 3D expression can be used.
Interestingly, Eq.~(\ref{alpha}) is written in a form in which the factor of $1/n$ is cancelled out by the linear increase in $n$ of $\ln[I(T_1)/I(T_2)]$ resulting in a value of $\lambda_{HAS}$ that is essentially independent of the monolayer number $n$, even for $n>n_{sat}$ where the 3D Fermi density of states is applied.
Analysis of the data for thermal attenuation of the specular He atom diffraction peak for up to 10 layers of the alkali metals Li, Na, K, Rb and Cs deposited on various metal substrates produced values for $\lambda_{HAS}$ that are in quite reasonable agreement with known tabulated values for bulk crystals, and similarly for up to 25 ML of Pb deposited on Cu(111).\cite{JPCL2}

The success of the theory expressed in Eqs.~(\ref{Eq1}) and (\ref{alpha}) in obtaining values of $\lambda_{HAS}$ in systems of layer-by-layer growth leads naturally to the question of applying similar theory to obtain $\lambda$ for other types of layered compounds.  In particular, there are several chalcogenide compounds for which DW attenuation measurements of He atom diffraction have been reported.  
We show below that with appropriate interpretation of Eqs.~(\ref{Eq1}) and (\ref{alpha}) reliable values of $\lambda_{HAS}$ can be extracted from such experimental data.

\subsection{Mass enhancement factor for a $d$-dimensional electron gas}  \label{d-dimension}


Dimensionality\cite{Coxeter} enters the expression of $\lambda_{HAS}$ through the Fermi-level
DOS per unit energy $\mathcal{N}_F^{(d)}$ and unit $d$-dimensional hyper-volume
\begin{eqnarray} \label{d1}
\mathcal{N}_F^{(d)} ~=~ \frac{k_F^d}{\gamma_d E_F}
~,
\end{eqnarray}
where
\begin{eqnarray} \label{d2}
\gamma_d  ~\equiv~ 2^{d-1} ~ \pi^{d/2} ~ \Gamma\left(\frac{d}{2}\right)
~,
\end{eqnarray}
$d$ being the dimension,
and $\Gamma$ the Riemann gamma-function, which for integer $d$ has values given by
\begin{eqnarray} \label{d3}
\Gamma\left(\frac{d}{2}\right) ~=~  \left(\frac{d}{2} -1 \right) ! ~;  ~~~~~\mbox{d even},
\nonumber
\\
~=~ \frac{(d-2)!! \sqrt{\pi}}{2^{(d-1)/2}} ~; ~~~~\mbox{d odd}.
\end{eqnarray}
For $d=2$, $\gamma_2 = 2 \pi$, and $d=3$, $\gamma_3 = 2 \pi^2$, the  usual  two-dimensional electron gas (2DEG) and three-dimensional electron gas (3DEG) 
expressions, namely $\mathcal{N}_F^{(2)} = m_e^*/ \pi \hbar^2$ and
$\mathcal{N}_F^{(3)} =  m_e^* k_F / \pi^2 \hbar^2 $, respectively, are readily obtained with
$k_F = (2 m_e^* E_F)^{1/2}/\hbar$.
As discussed elsewhere,\cite{JPCL2} the 3DEG of a thick anisotropic degenerate semiconductor slab can be viewed as a stack  of a number $n = n_{sat}$ of 2DEGs. This yields a definition of $n_{sat}$ as
\begin{eqnarray} \label{2d3}
n_{sat}  ~=~ c^* \frac{k_{F\perp}}{\pi}
~,
\end{eqnarray}
where $k_{F\perp}$ is the Fermi wave vector normal to the surface and
$c^*$ has the meaning of the e-ph interaction range normal to the surface or the maximum depth beneath the surface from where phonon displacements can modulate the surface charge density. 
In this way, the 2D expression for the e-ph coupling constant for a thick layer crystal as given in Ref.~[\cite{JPCL2}] is just that of Eq.(\ref{alpha}) with $n=n_{sat}$.

For a general $d$-dimensional free-electron system (for any $d$, even fractional), one finds
\begin{eqnarray} \label{d5}
\lambda_{HAS}^{(d)} ~=~ - \frac{\phi \gamma_d}{4(k_F r_0)^d} ~\frac{k_F^2}{k_{iz}^2} ~
\frac{\partial \ln \{ I\left( T\right)\} }{k_B~ \partial T}
~,
\end{eqnarray}
where $r_0^d$ is the unit cell hyper-volume.
Similarly, in the case of a measurement of the dependence on the HAS specular reflectivity as a function of the incident wave vector at constant $T$, the expression for $\lambda_{HAS}^{(d)}$ becomes\cite{Manson-SurfSciRep}
\begin{eqnarray} \label{d6}
\lambda_{HAS}^{(d)} ~=~ - \frac{\phi \gamma_d}{4(k_F r_0)^d} ~\frac{k_F^2}{k_B T} ~
\frac{\partial \ln \left\{k_i^\eta I(T)\right\}}{\partial \left( k_{iz}^2 \right)}
~.
\end{eqnarray}
The factor of $k_i^\eta$ multiplying the intensity is a correction for the energy dependence of the incident beam flux.
The standard theoretical treatment of a jet beam nozzle expansion flow shows that the beam energy varies inversely as the square root of the stagnation temperature. The correction factor is then simply $k_i$, or $\eta = 1$,\cite{Beijerinck} although in some cases different behaviors on incident energy have been measured.
Further discussion on the dependence of the incident beam on stagnation temperature and pressure has been reported by Palau {\em et al}.\cite{Palau}

When dealing with layered semimetal surfaces, the free electron gas is protected by an anion surface layer leading to an essentially hard-wall potential plus a more or less deep attractive van der Waals potential, $k^2_{iz}$ should be corrected due to the presence of the attractive well before being repelled by the hard wall (Beeby correction\cite{Beeby}). This implies that both the incident and final normal momenta should  be replaced by
$k_z^2 \longrightarrow k_z^2 + 2mD/\hbar^2$, where $D$ is the attractive potential depth. Usually, the incident energy $E_i$  is generally much larger than $D$, so that this correction can be neglected. 
With regard to Eq.~(\ref{d6}), notice that the Beeby correction cancels out in the differential in the denominator, but retains a minor effect through the term  in the numerator. 

\section{One dimensional electron gas}  \label{one}

The high sensitivity of the HAS technique  permits the detection of weak surface charge density waves (CDW) that are difficult to detect with other surface techniques.  Recently, it has been shown that from the temperature dependence of the CDW diffraction peaks information about the e-ph interaction sustaining the CDW transition is possible.\cite{JPCL-2020} There is an instability below a critical temperature $T_c$  generally induced by the e-ph coupling according to the theory developed by Fr\"ohlich-Peierls\cite{Peierls,Frolich} or the Kelly-Falicov multivalley mechanism.\cite{Falicov-1,Falicov-2,Falicov-3} 
In the latter case, the phonon-induced transitions between narrow pockets (nests) literally realize what is meant as perfect nesting. 
The occurrence of a CDW instability below $T_c$  yields additional $T$-dependent diffraction peaks in the elastic scattering angular distribution at parallel wave vector transfers $\Delta {\bf K}$ equal or close to the nesting vectors $Q_c$ (e.g., $Q_c = 2k_F$ for the 1D Peierls mechanism).

When examining the thermal attenuation for a given diffraction peak intensity due to the DW factor, 
the corresponding wave vector $\Delta {\bf K}$ transfer parallel to the surface  equals to either a ${\bf G}$-vector of the normal surface lattice $(\Delta {\bf K}  = {\bf G}) $, or to a CDW wave vector $Q_c$. 
In this case,
Eq.~(\ref{d5}) can also be applied to diffraction peaks by simply replacing $4k_{iz}^2$ by $\Delta k_{z}^2 + \Delta {\bf K}^2$. Usually in HAS experiments, the condition
$\Delta {\bf K}^2 << \Delta k_{z}^2$
holds, therefore the $T$-dependence of the diffraction and specular peaks leads to a $\lambda_{HAS}$ value which is independent of the diffraction channel.
In Eq.~(\ref{Eq1-2}), the temperature dependence of $I(T)$ comes from thermal vibrations. However, this is no longer completely true when considering the diffraction from a surface CDW which forms below $T_c$ from a Fermi surface instability. Clearly, the temperature-dependent population of electron states near the Fermi level follows Fermi statistics.  Here, $I_0$ has an implicit dependence on $T$, which is generally negligible with respect to that of $W(T)$, except near $T_c$; in this case, its square root $\sqrt{I_0}$ can be considered as an order parameter,\cite{order,order-2} and vanishes when $T \longrightarrow T_c$ as $(1-T/T_c)^\beta$,
where $\beta$  is the order-parameter critical exponent (typically $\beta = 1/3$).\cite{Liu,GBdraft,Old-2H-Ta-paper} In the following, 1DEG examples are shown that, away from the critical region, a CDW diffraction peak may be used to extract $\lambda_{HAS}$.

\subsection{Bi(114)}  \label{Bi114}

The case of Bi(114), a topological surface exhibiting properties of a 1D free electron gas, as well as a CDW, has been discussed in an earlier letter.\cite{JPCL-2020} As shown in that letter, the weak 2D character of this surface, characterized by a long period (28.4 \AA) in the
$\overline{\mbox{X}} \, \overline{\mbox{Y}}$ direction, is responsible for the charge density wave, observed with HAS below $T_c \approx 280$ K, via the Kelly-Falicov multivalley $\overline{\mbox{X}}$ - $\overline{\mbox{X}}$ nesting mechanism.  The pronounced 1D metal character of Bi(114) \cite{Wells-2009} is confirmed by the reliable value $\lambda^{(1D)}_{HAS} = 0.45$ obtained by treating it as a 
1D system. 

\subsection{W(110):H($1\times1$): 1D versus 2D}  \label{W(110)}

\begin{figure*}
	\includegraphics[height=10cm]{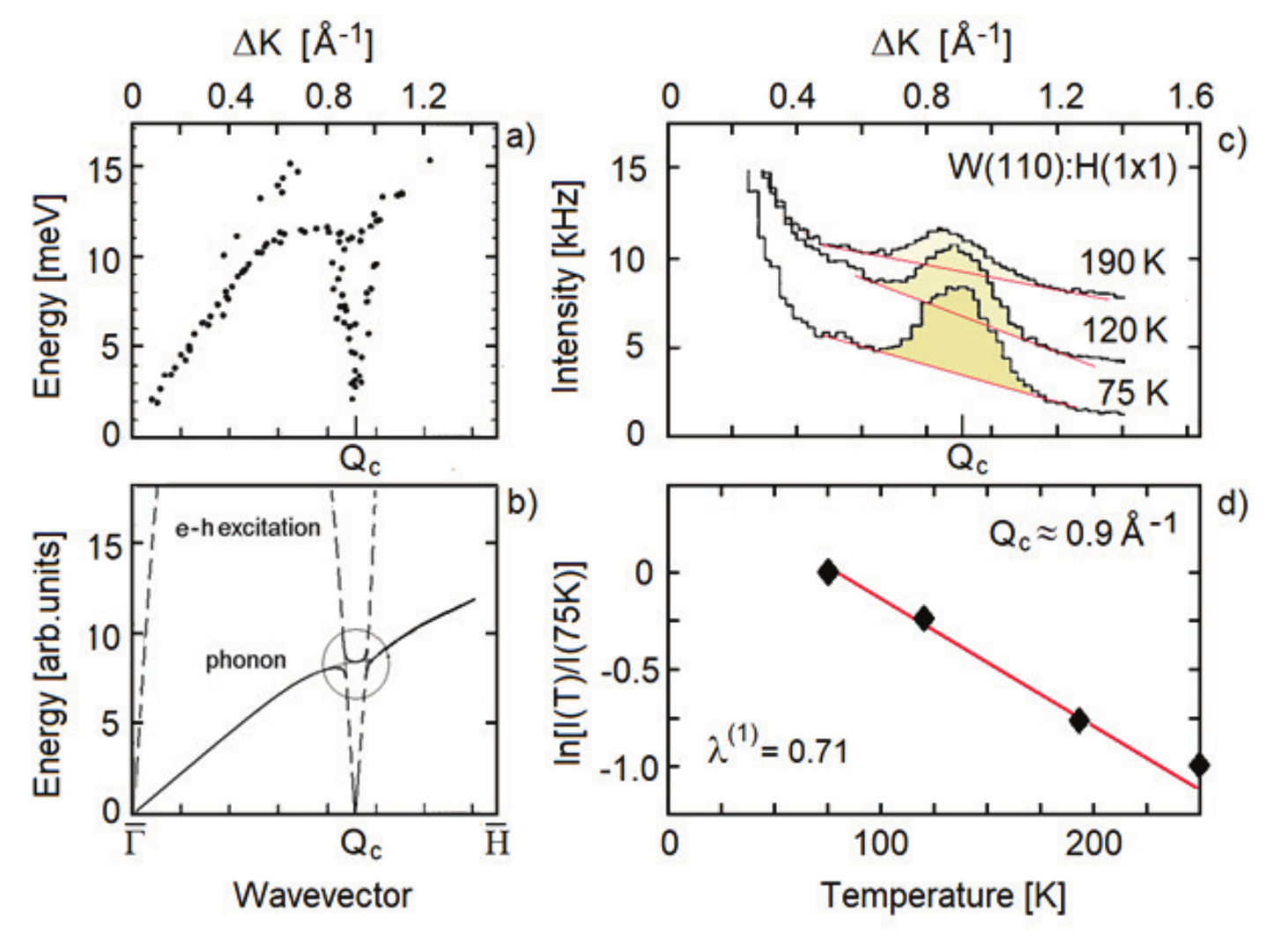}
	\caption{Helium atom scattering data from a hydrogen monolayer covering a tungsten (110) surface: W(110):H(1$\times$1).\cite{GBdraft}
		a) The giant anomaly observed in the Raliegh wave (RW) dispersion curve  at T = 130 K \cite{Hul93,Hul93-2} originates from an avoided crossing [as depicted in b)] with the electron-hole  excitation spectrum of a 1DEG associated with the  H(1$\times$1) overlayer. Ab-initio calculations by Kohler {\em et al.}\cite{Koh97}  indicate a Fermi-contour nesting vector $2k_F$ equal to the anomaly wave vector at $Q_c$. c) A HAS satellite diffraction peak at $Q_c$, suggesting the presence of a CDW, decreases with temperature with a linear dependence of the DW exponent as shown in d). Deviations from linearity beyond $T \sim 250$ K are attributed to coexistence with a disordered phase\cite{Gon83}  and desorption around 400 K.\cite{Hul93-2} The 1DEG e-ph coupling strength as derived from Eq.~(\ref{d5}) for dimensionality $d = 1$ is $\lambda_{HAS}^{(1D)} =0.71$.
	}
	\label{FigW110}
\end{figure*}

The H-saturated (110) surfaces of tungsten and molybdenum, despite the uniform distribution of the hydrogen atoms, which sit on hollow sites and form a lattice like that of the metal atoms, actually exhibit a pronounced 1DEG behavior along certain directions of the surface Brillouin zone (BZ) due to the favorable shape of surface state Fermi contours that provide good nesting conditions.\cite{Koh97} 
Such a 1D character of the surface electron gas has become evident from the observation with HAS of a giant sharp Kohn anomaly in the Rayleigh wave dispersion curve around the
$\overline{\Gamma}\, \overline{\mbox{H}}$
and  $\overline{\Gamma} \, \overline{\mbox{S}}$ directions, which are $[001]$ and $[1 \overline{1} 2]$ in the direct space, respectively, in both W(110):H(1$\times$1) [Fig.~\ref{FigW110}a)] and Mo(110):H(1$\times$1).\cite{Hul93-2} As schematically represented in Fig.~\ref{FigW110}b), the anomaly actually originates from the avoided crossing (encircled region) of the 1D electron-hole (e-h) excitation curve (broken line) with the phonon dispersion curve. 
As the increasing surface wave vector $\Delta {\bf K}$ approaches the anomaly wave vector $Q_c$, the e-ph coupling gradually turns the phonon excitation (where atoms move and electrons follow adiabatically) into an e-h excitation (where the charge density oscillates while dragging the atoms less and less), and then reverts back
to predominantly adiabatic phonon behavior as $\Delta {\bf K}$ continues to increase beyond $Q_c$. 
Inside the small encircled region around $Q_c$, i.e., in the vicinity of the anomalous behavior, the e-h excitation of the electron gas behaves like a phonon.
Remember that He atoms are scattered by the surface charge density,
and therefore exchange energy and momentum with the adiabatic phonons associated with the movements of the crystal cores, or with the anomalous (non-adiabatic) low-energy excitations of the surface nearly free-electron gas.
The presence of an approximately 2 meV gap in the e-h branch at $Q_c$ (although not much larger than the 1\% energy resolution of the incident energy $E_i = 34.8$ meV) suggests critical fluctuations of an incipient CDW, consistently with the observation of a small satellite peak at $Q_c  \cong 0.93$~\AA$^{-1}$ in the HAS diffraction spectrum 
along $\overline{\Gamma} \, \overline{\mbox{H}}$
[Fig.~\ref{FigW110}c)].\cite{Hul93,Lud94} The correspondence of the critical wave vector $Q_c$ to a nesting on the Fermi contour lines at $Q_c = 2k_F$, as resulting from photo-emission data and first-principle calculations, has been thoroughly discussed for both W(110):H(1$\times$1) and Mo(110):H(1$\times$1) by Kohler {\em et al.}\cite{Koh97}  The corresponding DW exponent decreases linearly with temperature [Fig.~\ref{FigW110}d)], at least in the temperature region $T \leq 225$ K, beyond which the ordered H(1$\times$1) phase starts coexisting with a disordered phase.\cite{Gon83}
The e-ph coupling constant $\lambda_{HAS}$  for W(110):H(1$\times$1) can now be extracted from the linear slope of the DW exponent. With the input data $\phi = 4.72$ eV for 
W(110):H(1$\times$1)\cite{Bar74}, $r_0 = \pi/k_F = 6.98$~\AA, and $(\Delta k_z)^2 + (2 k_F)^2 = 108.9$~\AA$^{-2}$ one obtains $ \lambda_{HAS}^{(1D)}   = 0.71$. This value compares well with the Rotenberg and Kevan result $\lambda = 0.8 \pm 0.2$, as derived from ARPES data collected along an azimuth close to the 
$\overline{\Gamma}\,\overline{\mbox{H}}$ direction (slice C in 
Figs.~1 and~2 of their cited paper).\cite{Rot02}

Unlike the CDW appearing in the case of Bi(114),\cite{JPCL-2020}
where the 1D character of the surface electron gas is quite evident from the large separation of the surface atomic chains, in the case of W(110):H(1$\times$1) the hydrogen, atoms form a 2D lattice like that of the clean W(110) surface. Moreover, a similar symmetry is expected for the CDW from the calculated map of the local probability function for e-h excitations across the surface state Fermi contours.\cite{Koh97} It is therefore plausible to derive $\lambda_{HAS}^{(2D)}$ from the same DW input data and compare it to  $\lambda_{HAS}^{(1D)}$. From the ratio		                        	
\begin{eqnarray} \label{d8}
\frac{\lambda_{HAS}^{(2D)}}{\lambda_{HAS}^{(1D)}} ~=~ \frac{\gamma_2}{\gamma_1}~\frac{r_0}{k_F a_c}
~,
\end{eqnarray}
where ${\gamma_2/ \gamma_1} = 2$ and $a_c = r_0^2/\sqrt{2}$  is the unit cell area for the CDW lattice. This gives ${\lambda_{HAS}^{(2D)}}/{\lambda_{HAS}^{(1D)}} = 2^{3/2}/\pi = 0.90$ , which means that also the 2D expression provides a reasonable approximation for W(110):H(1$\times$1). Since Eqs.~(\ref{d5}) and~(\ref{d6}) have been derived for an isotropic nDEG, the deviation of the above ratio from unity essentially measures the deviation from isotropy of the actual 2DEG. Genuine 2DEG systems are now considered in the next Section.

\section{2D topological materials. Chalcogenides}  \label{2D}

\begin{figure}[h]
	\includegraphics[height=10cm]{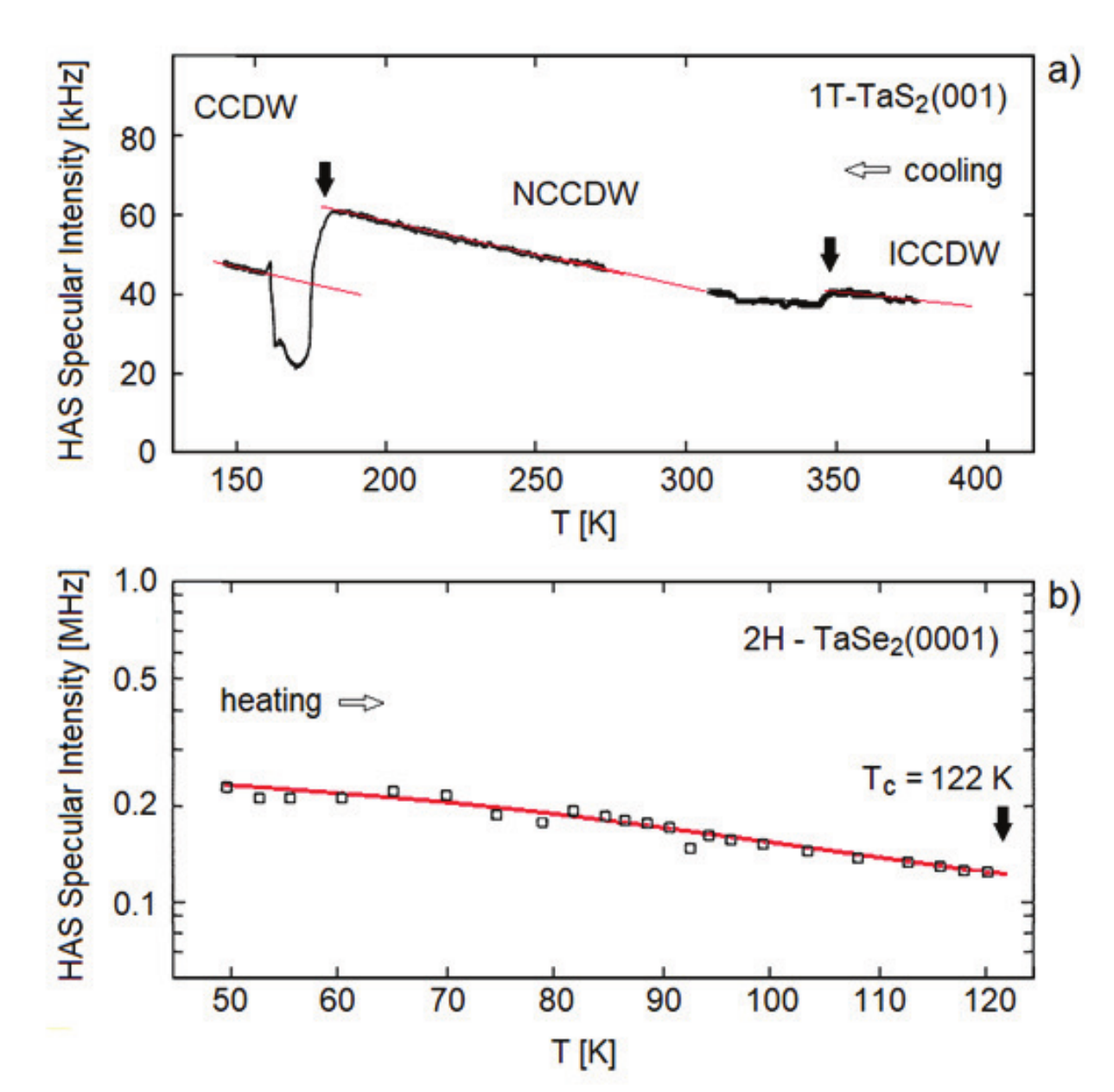}
	\caption{Thermal attenuation (Debye-Waller) data for scattering of He atoms from two chalcogenide surfaces. The specular intensity is plotted as a function of temperature.  a) 1T-TaS$_2$(001) plotted on a linear scale, and b) 2H-TaSe$_2$(0001) plotted on a logarithmic scale.  
		In a) the crystal was initially at high temperature and then cooled, and
		the two vertical arrows indicate the charge density wave transitions.
		The transition from incommensurate charge density wave (ICCDW) to non-commensurate occurs at about 350 K, while
		the transition from non-commensurate  to commensurate charge density wave (CCDW) occurs at about 180 K.
		In b) the temperature was increased starting from the smallest $T$ and the charge  density wave transition occurs a 122 K as marked by the arrow.  Data are from Ref.~[\cite{Heimlich}].
	}
	\label{FigTaS2-TaSe2}
\end{figure}

The cases of recently measured topological Bi pnictogens Bi$_2$Se$_3$,  Bi$_2$Te$_3$ and Bi$_2$Te$_2$Se  have been considered elsewhere.\cite{JPCL-2020}
%
\begin{table*}
	\caption{
		The mass enhancement factor $\lambda_{HAS}$ for some transition-metal chalcogenides 
		as determined from the temperature dependence of the thermal attenuation of elastic specular He atom  diffraction. 
		The entry for PtTe$_2$
		was evaluated from the dependence on incident angle of the diffuse elastic peak intensity at a constant temperature of 100 K. The two different values of $\lambda_{TF}$ and $\lambda_{HAS}$ reported for MoS$_2$ correspond to two samples with surface carrier concentrations of 5 and 7 $10^{12}$ cm$^{-2}$, respectively. \cite{Ane}
		The  Beeby correction $D$  for  MoS$_2$  is $13.6$ meV. 
	}
	\vspace{1cm}
	\centering
	\begin{tabular}{|c||c|c|c|c|c|c|}
		\hline      Surface   & $T$ range
		& $ k_i^2$ & $ \phi $  & $\lambda_{TF}$ & $\lambda_{HAS}$ & $\lambda$     \\
		& $[K]$  & [\AA$^{-2}$]& $\left[ \mbox{eV} \right]$ & \AA & & (other sources)  \\
		\hline
		%
		%
		%
		%
		%
		%
		%
		\hline 2H-MoS$_2$(001)     \cite{Ane}   &95-450 &  121  &
		5.2 \cite{Cho} & $ 9.8$ \cite{c-Pee}  &   0.41    & $\sim$0.1  \cite{LiC,Ge} \\
		& & & &9.3 &0.49 &0.12-0.20 \cite{Nay} \\
		\hline 1T-TaS$_2$(001)     \cite{Heimlich}   &180-280 \cite{Yu,a-Yu} &  29.2  &
		5.2 \cite{Shi} & $10.2$ \cite{d-Yu} & 1.0 \cite{a-Yu}   & 1.0 \cite{Ros} \\
		&350-380 \cite{b-Yu} & & & &$\sim$0.4 \cite{b-Yu} &0.69-2.09 \cite{Liu} \\
		& & & & & & 0.38 \cite{Hin} \\
		\hline 2H-TaSe$_2$     \cite{Heimlich}   &50-120 &  137  &
		5.5 \cite{Tso} & 12.6 \cite{NTaSe2} & 0.58   & 0.49  \cite{Ros} \\
		& & & & & &0.39 \cite{Bho} \\
		\hline 1T-PdTe$_2$(001)     \cite{Far2}   &50-300 &  121  &
		4.6 \cite{Ras} & 10.62 \cite{Far2} & 0.58    & 0.59  \cite{Hoo} \\
		& & & & & &0.53 \cite{Kim} \\
		%
		%
		%
		\hline 1T-PtTe$_2$   \cite{Far1}   &100 & 159   & 4.52 \cite{Ras}
		& 10.4\cite{Far1} & 0.40 $\pm$ 0.02     & 0.35 \cite{Kim} \\
		\hline
	\end{tabular}
	\label{chalcogenides}
\end{table*}
%
In addition, several other chalcogenide crystal surfaces that have been investigated with He atom scattering are listed in
Table~\ref{chalcogenides}.  In all but one of these systems it was the specular thermal attenuation that was measured
and two examples, for 1T-TaS$_2$ and 2H-TaSe$_2$, are shown in Fig.~\ref{FigTaS2-TaSe2}.
Although the experiments on the chalcogenides did not involve layer-by-layer growth we can use for their analysis Eq.~(\ref{Eq1}) together with the definition of $\alpha$ of Eq.(\ref{alpha}) which leads to
\begin{eqnarray} \label{C1}
\lambda_{HAS} ~=~
\frac{\pi}{2 n_s}~ \alpha
~,
\end{eqnarray}
with $n_s$ given by
\begin{eqnarray} \label{C2}
n_s ~=~ \frac{\pi \hbar^2 \mathcal{N}(E_F)}{m_e^* a_c}
~.
\end{eqnarray}
The quantity $n_s$ plays a role similar to $n_{sat}$ in the case of layer-by-layer growth, i.e., it is the number of layers whose electronic states at the Fermi level concur to give the surface charge density probed by the He atoms, and $\mathcal {N} (E_F)$ is their total density of states; thus $\mathcal{N}(E_F)= n_s a_c \mathcal{N}^{(2)}_F$. This implies that $n_s$ represents the number of layers of the compound that are contributing to the value of $\lambda_{HAS}$ as measured in a He atom collision with the surface.
Eqs.~(\ref{C1}) and~(\ref{C2}) suggest that the question of determining the density of states $\mathcal{N}(E_F)$ appropriate to the reflection of atomic He is cast into the problem of determining a single parameter, namely the small number of layers $n_s$ of the surface that contribute.

%
%
%
%
%
%
%

In semimetals and degenerate semiconductors, as well as in normal semiconductors with a degenerate accumulation layer at the surface, the density of states $\mathcal{N}(E_F)$ is associated with the presence of surface states and quantum-wells states confined within the band bending, i.e., within the Thomas-Fermi (TF) screening length $\lambda_{TF}$. \cite{Efros-Sklovskii-book} 
In transition-metal layered compounds the limited electron mobility in the normal direction, resulting in a large effective mass anisotropy, makes $\lambda_{TF}$ comparatively short, typically spanning about two triple layers. Thus $n_s \approx 2$ appears to be an appropriate value for this class of materials (see Table~\ref{chalcogenides}). 
Pnictogen chalcogenides, whose e-ph coupling strength as derived from HAS measurements have been presented and discussed elsewhere, \cite{JPCL-2020} are characterized by screening lengths an order of magnitude larger than in transition metal chalcogenides, due to their quintuple layer structure with a more pronounced 3D character. As a consequence, their $\lambda_{HAS}$ is seen to receive a far larger contribution from the surface quantum-well states than from the topological Dirac states. 

Table~\ref{chalcogenides} lists some chalcogenides, 
not including the Bi pnictogens considered in Ref.~[\cite{JPCL-2020}],
for which sufficient temperature-dependent data are available for He atom scattering, together with the corresponding references, the relevant experimental parameters, and the values obtained for $\lambda_{TF}$.
In each case shown in Table~\ref{chalcogenides}, with reasonable choices for $\lambda_{TF}$ determined using parameters taken from the literature, results for $\lambda_{HAS}$ are in fairly good agreement with values of $\lambda$ determined from other known sources, which may be either bulk measurements or calculations.
The entry for PtTe$_2$  differs from the others in that the elastic peak measured was not the specular one but  the off-specular diffuse elastic peak measured at the constant temperature of 100 K. In this case, the incident angle was varied, keeping the source-to-detector angle fixed. Thus,  Eq.~(\ref{alpha}) was modified to account for the correct wave vector difference, i.e.,  $k_{iz}^2 \rightarrow \Delta {\bf k}^2/4 = ({\bf k_f} - {\bf k_i})^2/4$, and the factor $\alpha$  changes to reflect the fact that the intensities were evaluated at different incident angles, but at constant $T$.

\section{Quasicrystalline surfaces}  \label{quasi}

Quasicrystals (QC) are characterized by a long-range orientational order \cite{She84,Sta99,Suc02} but no periodicity. 
They can be viewed as projections onto the ordinary space of a periodic lattice in a space $n$D of higher dimension. \cite{Sad90} Certain 
structural and dynamical properties of the QC may be more conveniently described in the corresponding $n$D periodic lattice. This representation 
is adopted here to derive the e-ph coupling constant for two QC metallic structures, a dodecagonal bilayer and a 3D decagonal QC, respectively,
represented by periodic 4D and 5D lattices.

\subsection{2ML-Ba(0001)/Cu(001)}  \label{Ba}

\begin{figure}[h]
	\includegraphics[height=10cm]{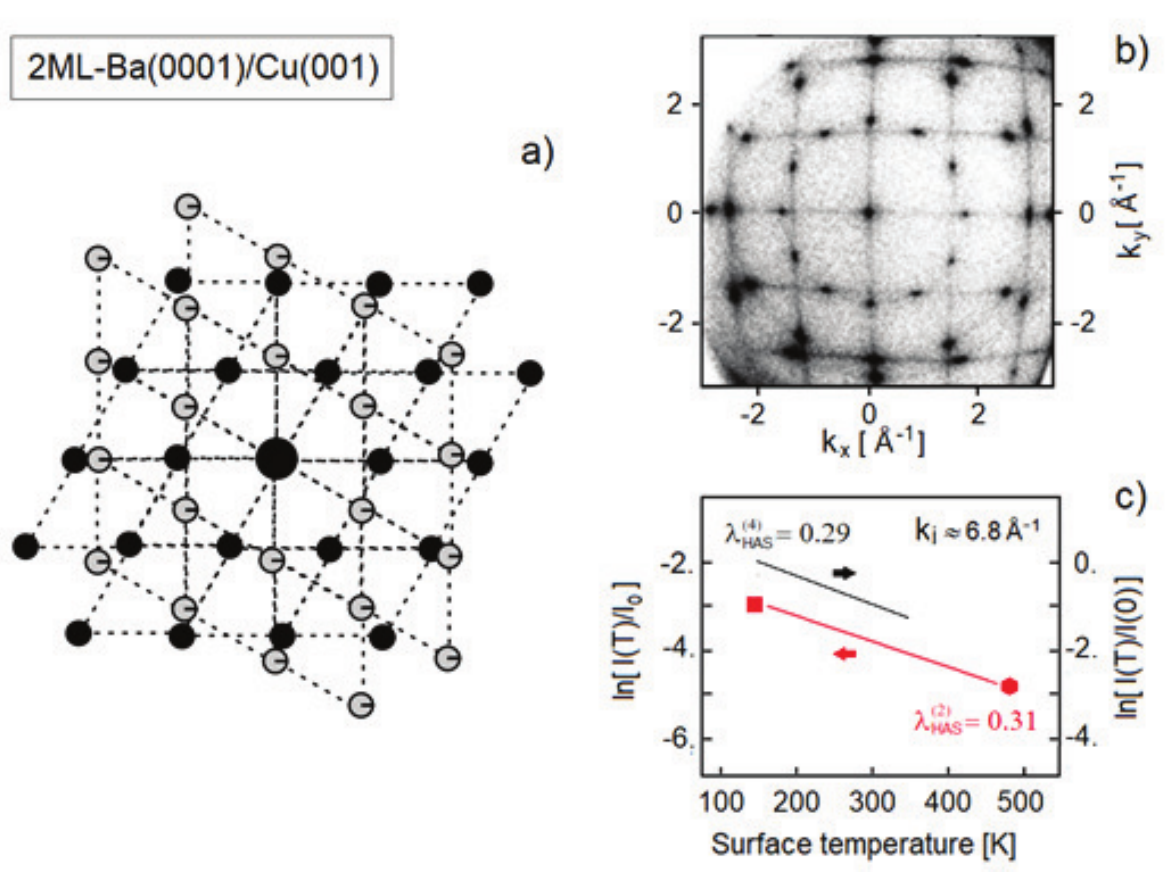}
	\caption{The surface of two monolayers of Ba(0001) on Cu(001).
		a) A Ba bilayer grown on Cu(001) at growth stage III \cite{Bar01} exhibits a dodecagonal QC structure corresponding to the LEED pattern shown in 
		b) The structure is a superposition of two hexagonal 2D lattices with a 30$^{\circ}$ twist, and is the projection of a 4D $\{3,3,4,3\}$ honeycomb 
		lattice \cite{Sad90}. c) The HAS reflectivity DW exponent measured at two different temperatures and the same incident wavevector $k_i = 6.8$ 
		$\mbox{\AA}^{-1}$ allows derivation of the e-ph coupling constant by either treating the bilayer as a 
		2D ($\lambda^{(2D)}_{HAS}$) or a 4D  ($\lambda^{(4D)}_{HAS}$) system, with little difference between the two results. 
		The latter is closer to the current value for bulk barium ($\lambda = 0.27$) as reported by Allen. \cite{Allen} }
	\label{FigBa0001}
\end{figure}

The LEED pattern for a barium bilayer grown on Cu(001) at the growth stage denoted  as III by Bortholmei {\em et al.} \cite{Bar01} exhibits a dodecagonal 
QC structure as shown in Fig. \ref{FigBa0001}a,b).  The structure is a superposition of two hexagonal 2D lattices with a $30^{\circ}$ twist. A twist of any angle other than 
integer multiples of $\pi/3$ 
breaks the hexagonal periodicity of a single layer while keeping a long-range orientational order. The present structure can be generated by the projection of a 4D $\{3,3,4,3\}$ honeycomb lattice (actually a 4D-bcc cubic lattice) \cite{Sad90}. 

The DW exponent
measured from the HAS reflectivity $I(T)$ as a function of temperature at a given incident wave vector $k_i \approx 6.8$ $\mbox{\AA}^{-1}$ as shown in Fig.~\ref{FigBa0001}c),
\cite{Bar01} allows the derivation the e-ph coupling constant from Eq. (\ref{d5}) for dimensionality $d = 4$ using two different estimates. First, the temperature dependence of $2W(T)$ has been obtained using 
the Debye temperature $\Theta_D = 120$ K reported in Ref.~[\cite{Bar01}] for the stage III Ba bilayer in the temperature range $150-350$ K using
the prescription of Ref.~[\cite{Ben18}], where $M$ is the Ba atom mass. This gives a slope of $77.2$ $eV^{-1}$ 
[Fig. \ref{FigBa0001} c), black line]. 
A second method is the direct 
derivation of the slope from reflectivity at two different temperatures, 480 K and 145 K, which gives a similar value of 67.9 $eV^{-1}$ 
[Fig. \ref{FigBa0001} c), 
red line connecting the two data points], although the low-temperature bilayer structure shows a pronounced four-fold symmetry, compatible with the 2D periodicity of 
the bcc-Ba(001) surface.\cite{Bar01} 
With the slope of 77.2 $eV^{-1}$, $\gamma_4 = 8 \pi^2$,  taking as a simple estimate $\phi = 2.7 eV$  \cite{k} and 
$k_F = 0.28$ $\mbox{\AA}^{-1}$ for Ba metal, \cite{Kit05} and $r_0$ 
(where $r_0^4$ gives  the 4D-bcc hypercell volume) equal to the lattice parameter 
of bcc Ba (5.03 $\mbox{\AA}$), it is found that $\lambda^{(4D)}_{HAS} = 0.29$. For comparison, the slope of 67.9 eV$^{-1}$ used with the 2D version 
of Eq. (\ref{d6}) gives  $\lambda^{(2D)}_{HAS} = 0.31$. The value of $\lambda^{(4D)}_{HAS}$  is closer to the current value for bulk barium 
( $\lambda= 0.27$) as reported in Ref.~[\cite{Allen}], but the difference is not particularly significant at the present level of approximation.

\begin{figure}[h]
	\includegraphics[height=10cm]{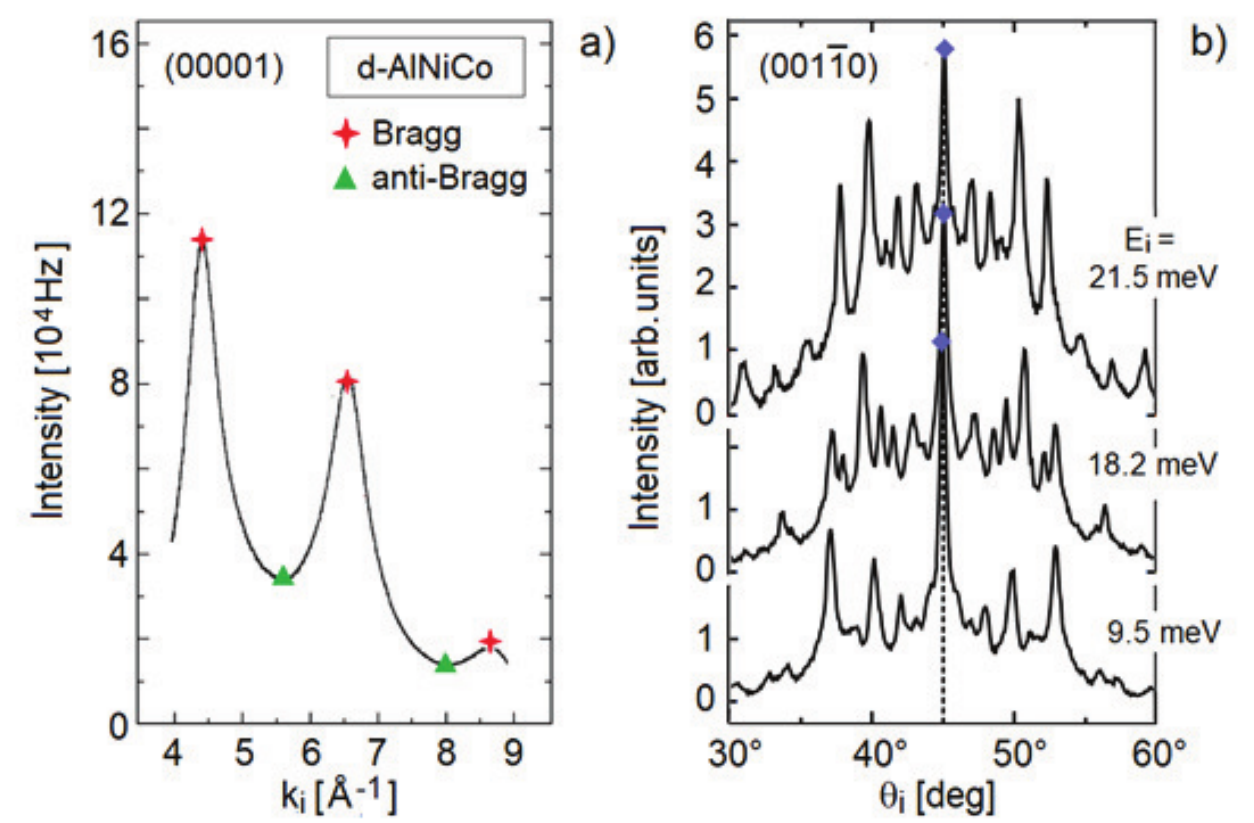}
	\caption{
		a) HAS reflectivity of the d-AlNiCo(00001) surface measured by Sharma {\em et al.} \cite{Sha04} as function of the incident wave vector at room temperature, with the scattering plane along the symmetry direction [10000]. The large oscillations are due to constructive interference (Bragg, marked with red stars) or destructive (anti-Bragg, marked with green triangles) interference at surface steps. b) HAS diffraction patterns from the d-AlNiCo  $(001\bar{1}0)$ surface measured at different incident energies and room temperature along the [10000] direction. 
		The specular peak (blue diamonds) intensities are derived after subtraction of the multiphonon Gaussian background. The incident energies fulfill approximately Bragg conditions (adapted from 
		Sharma {\em et al.}\cite{Sha02}). }
	\label{Fig5DQC}
\end{figure}

\subsection{d-AlNiCo(00001)}  \label{AlNiCo}

The surface structure and dynamics of the decagonal quasicrystal Al$_{71.8}$Ni$_{14.8}$Co$_{13.4}$ (approximately Al$_5$NiCo), hereafter termed d-AlNiCo, have been investigated by HAS for the surfaces (00001) \cite{Sha03,Sha04,The04} and ($001\overline{1}0$) \cite{Sha02} of the nominal 5D generating periodic lattice. \cite{Sad90} The available data permit determination of information on the dependence of the specular intensity on the incident momentum  and hence evaluation of the e-ph coupling constant by using Eq.~(\ref{d6}). 

\begin{figure*}
	\includegraphics[height=6cm]{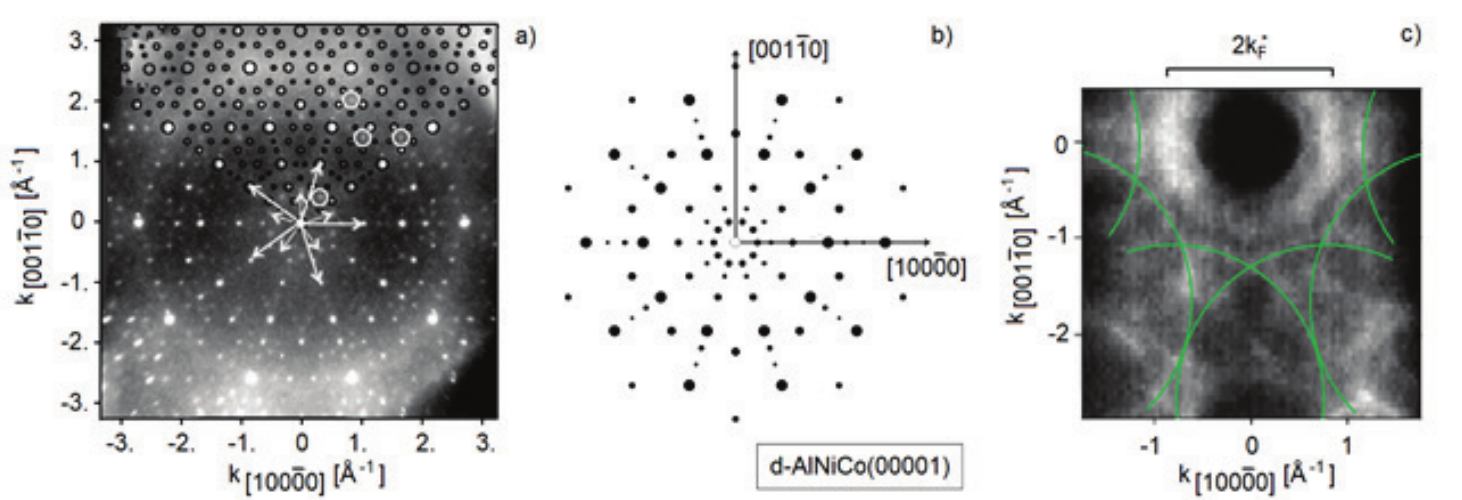}
	\caption{
		a) SPA-LEED image of d-AlNiCo(00001), with indication of the basis vectors (arrows) of b) the 5D reciprocal lattice 
		(reproduced from Ref.~[\cite{Sha04}]), 
		and c) a Fermi surface cut as imaged by SX-ARPES along a plane parallel to the surface at $9G_z$ ($G_z = 2 \pi/c$) (adapted from Ref.~[\cite{Rog15}]) with the indication at the top of panel c)
		of the diameter $2k_F = 1.8 \mbox{\AA}^{-1}$ of the most prominent Fermi contour. The green circles represent the Al $sp$-band Fermi contours of diameter $ 2 k_F^0 = 3.14~\mbox{\AA}^{-1}$. \cite{Rog15} }
	\label{Fig5DQC-2}
\end{figure*}

Figure~\ref{Fig5DQC}a) shows the HAS specular intensity for the (00001) surface of d-AlNiCo reported by Sharma {\em et al.} \cite{Sha04} as a function of the incident wavevector in a 90$^\circ$ scattering geometry 
($ k_{iz} =  ki / \sqrt{2}$) at room temperature. The large oscillations are due to the interference between beams reflected above and below a step: the maxima (Bragg scattering) occur when $k_{iz} = \pi n/s$, 
with $s$ the step height and $n$ an integer; the minima (anti-Bragg scattering) where  $ k_{iz} = \pi (n+1/2)/s$ (here $n = 2,~3,~4$). The step height corresponding to the peak positions $s = 2.06$~\AA~\cite{Sha04} is that of a monolayer, in agreement with the bilayer periodicity of $\sim 4$~\AA~reported by Rogalev {\em et al.} \cite{Rog15} The average slopes of the Bragg and anti-Bragg peak intensities as functions of $k_{iz}$ are
$- \partial\ln(I/k_i)/\partial(k_{iz}^2) = 0.086 \pm 0.017 \mbox{\AA}^{-2}$ (Bragg)
and 0.056~\AA$^{-2}$ (anti-Bragg).
The difference between the two slopes indicates an appreciable contribution of the steps to the e-ph coupling, which is present in Bragg scattering and less so in anti-Bragg scattering.
The HAS diffraction patterns have also been measured by Sharma {\em et al.} \cite{Sha02} for the d-AlNiCo  surface at three different incident energies along the quasi-periodic [10000] direction as shown in Fig.~\ref{Fig5DQC}b). The dependence on the incident wavevector of the specular elastic peak intensity, after subtraction of the multiphonon bell-shaped background, is given by
$- \partial\ln(I/k_i)/\partial(k_{iz}^2) = 0.080 \pm 0.021~\mbox{\AA}^{-2}$.

SX-ARPES data by Rogalev {\em et al.} \cite{Rog15} from d-AlNiCo(00001) provide information on the projection of the 5D Fermi surface onto various cuts of the 3D reciprocal space like the one, reproduced in Fig.~\ref{Fig5DQC-2}c),  taken parallel to the ($x,y$) 
plane at $k_z = 9(2\pi/c)$ 
where $c = 4$~\AA~ is the surface bilayer thickness, periodically repeated in the z direction [00001]. As it appears from this cut of SX-ARPES data, $k_F = 0.9~\mbox{\AA}^{-1}$ seems to be a reasonable choice (Fig.~\ref{Fig5DQC-2}c)). Then, with $\phi= 4.8$ eV from He$^*(2^3S,1s2s)$ de-excitation spectroscopy, \cite{Suz05} $r_0 \cong 6.0$~\AA~from the value of the reciprocal quasi-lattice vector $G = 1.04~\mbox{\AA}^{-1}$ derived from the HAS dispersion curves in the [10000] direction \cite{Sha03} (but also from HREED data \cite{Suz05}) and $k_B T = 0.026$ eV, Eq.~(\ref{d6}) gives
$\lambda^{(5D)}_{HAS} = 0.26 \pm 0.05$ (Bragg) and 0.17 (anti-Bragg)
for the d-AlNiCo(00001) surface.
As argued above, the larger value of $\lambda^{(5D)}_{HAS}$   obtained from Bragg reflectivity may be attributed to an enhancement of the e-ph interaction at steps.  With the HAS data for the 
($001\overline{1} 0$)  surface and the same input parameters $k_F,~\phi,~r_0,~\mbox{and}~ k_BT$ given above one finds
$\lambda^{(5D)}_{HAS} = 0.24 \pm 0.06$.
The similarity of this value with that for the (00001) surface including steps can be understood from the fact that the ($001 \overline{1} 0$)  surface actually exhibits (00001) facets \cite{Sha02}, which can {\em a posteriori} justify the use of the same (00001) input data for the derivation of $\lambda^{(5D)}_{HAS} $.

Finally, it is interesting to discuss the possibility that the e-ph interaction probed by HAS is actually restricted to the surface, so that Eq. (\ref{d6}) with dimension d = 4 is more appropriate. In this case the same input data yields slightly larger values for the e-ph coupling,
$\lambda^{(4D)}_{HAS} = 0.29 \pm 0.06$ (Bragg) and 0.20 (anti-Bragg)
for the d-AlNiCo(00001) surface,  and $0.22 \pm 0.06$ for the ($0 0 1 \overline{1} 0$).
Considering that also the HAS reflectivity data for the $(001\bar{1}0)$ surface are approximately under Bragg conditions, the comparison with the Bragg value of $\lambda_{HAS}^{(4D)}$ for the $(00001)$ surface indicates an appreciable anisotropy, which favors the quasicrystalline plane.
These values can be compared with the mass-enhancement factor extracted from the e-ph enhancement of thermoelectric power (TP) measurements in Y-AlNiCo, the monoclinic 3D approximant of decagonal d-AlNiCo. \cite{Dol09} At room temperature the fit in the $a^*$ direction (approximating [10000] in the decagonal sample) gives $\lambda_{TP} \sim 0.2$, a value which appears, however, to increase at lower temperature. \cite{Dol09}
There is significant anisotropy,
with $\lambda_{TP}$ for the quasicrystalline plane about five times larger than along the tenfold axis. Moreover, $\lambda_{TP}$ exhibits a large increase for decreasing $T$, though the present value $\lambda_{HAS}^{(4D)} = 0.28$ from room temperature HAS data would correspond to the Shuyuan {\em et al.} TP data at about 130 K.\cite{Shuy96}

It may be argued that an ideal free-electron model  does not contain information on the lattice periodicity or quasi-periodicity. 
With this in mind, the 5D d-AlNiCo can also be treated as a 3D system with a Fermi surface from the highly-dispersed 
Al $sp$-band ($ k_F^0 = 1.57~\mbox{\AA}^{-1}$). \cite{Rog15} In this case, with the same $\phi$  and $r_0$, 
one finds
$\lambda^{(3D)}_{HAS} = 0.23 \pm 0.05$ (Bragg) and 0.15 (anti-Bragg)
for the d=AlNiCo(00001) surface, and $0.22 \pm 0.06$ for the ($0 0 1 \overline{1} 0$).
It appears that, within the experimental uncertainties, the effects of dimensionality are in this case rather modest. On the other hand, it is easily seen that treating d-AlNiCo(00001) as a 2D system, with the response restricted to the surface bilayer ($n_{sat} = 2$), 
the ratio  
$\lambda^{(2D)}_{HAS}/\lambda^{(4D)}_{HAS} = (r_0 k_F)^2/4 \pi n_{sat}$
is for this case equal to 1.16. Thus, in this case, like for 2ML-Ba/Cu(001), the 2D treatment yields only a slightly larger e-ph coupling constant than that obtained from the treatment with the more appropriate dimensions.

\section{Conclusions}

In this paper, we have analyzed the thermal dependence of the Debye-Waller factor measured in the scattering of atoms from a selection of complex surface systems in order to extract values of the electron-phonon coupling constant $\lambda$.  The analysis is based on a theory originally developed for obtaining $\lambda$ for metal surfaces, but which here is adapted to the more complicated cases of layered chalcogenide semiconductors and systems that can be considered as having different dimensions such as 1D CDWs and quasicrystals.
The original theory demonstrates that the argument of the Debye-Waller factor $2W({\bf k}_f, {\bf k}_i,T)$
is, to a good approximation, proportional to $\lambda$.  The current analysis shows that, with suitable interpretation of the theory, values of $\lambda$ can be obtained from the surfaces of these 
more complex systems.
For all of these systems, the values of $\lambda_{HAS}$ obtained from atom-surface scattering experiments compare favorably with established values for the bulk materials as published in the literature.

~\\~\\
{\bf Acknowledgments}: One of us (GB) would  like to thank Prof.~Marco Bernasconi for helpful discussions.  This work is partially 
supported by a   grant with Ref. FIS2017-83473-C2-1-P from the Ministerio de Ciencia, Universidades e Innovaci\'on (Spain).


\end{document}